\documentclass[aps,pra,amsmath,twocolumn,superscriptaddress]{revtex4-2}
\usepackage{graphicx}
\usepackage{color}
\usepackage{xspace}

\begin{document}

\title{Effect of controlled artificial disorder on the magnetic properties of\\ EuFe$_2$(As$_{1-x}$P$_{x }$)$_2$ ferromagnetic superconductor}

\author{Sunil Ghimire}
\affiliation{Ames Laboratory, Ames, Iowa 50011, USA}
\affiliation{Department of Physics \& Astronomy, Iowa State University, Ames, Iowa 50011, USA}
	
\author{Marcin Kończykowski}
\affiliation{Laboratoire des Solides Irradiés, CEA/DRF/lRAMIS, École Polytechnique, CNRS, Institut Polytechnique de Paris, F-91128 Palaiseau, France}

\author{Kyuil Cho}
\affiliation{Ames Laboratory, Ames, Iowa 50011, USA}

\author{Makariy A. Tanatar}
\affiliation{Ames Laboratory, Ames, Iowa 50011, USA}
\affiliation{Department of Physics \& Astronomy, Iowa State University, Ames, Iowa 50011, USA}
	
\author{Daniele Torsello}
\affiliation{Politecnico di Torino, Department of Applied Science and Technology, Torino 10129, Italy}
	
\author{Ivan S. Veshchunov}
\affiliation{Department of Applied Physics, The University of Tokyo, Hongo, Bunkyo-ku, Tokyo 113-8656, Japan}
	
\author{Tsuyoshi Tamegai}
\affiliation{Department of Applied Physics, The University of Tokyo, Hongo, Bunkyo-ku, Tokyo 113-8656, Japan}
	
\author{Gianluca Ghigo}
\affiliation{Politecnico di Torino, Department of Applied Science and Technology, Torino 10129, Italy}
\affiliation{Istituto Nazionale di Fisica Nucleare, Sezione di Torino, Torino 10125, Italy}

\author{Ruslan Prozorov}
\email[Corresponding author: ]{prozorov@ameslab.gov}
\affiliation{Ames Laboratory, Ames, Iowa 50011, USA}
\affiliation{Department of Physics \& Astronomy, Iowa State University, Ames, Iowa 50011, USA}
	
	\date{\today}
	\begin{abstract}
		Static (DC) and dynamic (AC, at 14 MHz and 8 GHz) magnetic susceptibilities of single crystals
		of a ferromagnetic superconductor, $\textrm{EuFe}_{2}(\textrm{As}_{1-x}\textrm{P}_{x})_{2}$
		(x = 0.23), were measured in pristine state and after different doses
		of 2.5 MeV electron or 3.5 MeV proton irradiation. The superconducting transition temperature, $T_{c}(H)$, shows an extraordinarily large decrease. It starts at $T_{c}(H=0)\approx24\:\textrm{K}$
		in the pristine sample for both AC and DC measurements, but moves to almost half of that value
		after moderate irradiation dose. Remarkably, after the irradiation
		not only $T_{c}$ moves significantly below the FM transition, its
		values differ drastically for measurements at different frequencies, $\approx16$
		K in AC measurements and $\approx 12$ K in a DC regime. We attribute
		such a large difference in $T_{c}$ to the appearance of the spontaneous internal magnetic field below
		the FM transition, so that the superconductivity develops directly
		into the mixed spontaneous vortex - antivortex state where the onset of diamagnetism is known
		to be frequency-dependent. We also examined the response to the applied DC magnetic fields and studied the annealing of irradiated samples, which almost completely restores the superconducting transition. Overall,
		our results suggest that in $\textrm{EuFe}_{2}(\textrm{As}_{1-x}\textrm{P}_{x})_{2}$ superconductivity is affected by local-moment ferromagnetism mostly via the spontaneous internal magnetic fields induced
		by the FM subsystem. Another mechanism is revealed upon irradiation where magnetic defects created in ordered  $\text{Eu}^{2+}$ lattice act as efficient pairbreakers leading to a significant $T_{c}$ reduction upon irradiation compared to other 122 compounds. On the other hand, the exchange interactions seem to be weakly screened by the superconducting phase leading to a modest increase of $T_{m}$ (less than 1 K) after the irradiation drives $T_{c}$ to below $T_{m}$. Our results suggest that FM and SC phases coexist microscopically in the same volume.
	\end{abstract}
	
\maketitle

\section{Introduction}
Coexistence and competition of superconductivity and magnetism is
a fascinating and actively studied topic. It is impossible to give even
remotely-complete reference list, see for example Refs. \cite{Ginzburg1957,Matthias1958,Buzdin1984,Machida1984,Bulaevskii1985,Buzdin1986,Sinha1989,Fischer1990,Maple1995,Flouquet2002}.
Full local-moment ferromagnetism can destroy superconductivity even
when it forms well below the superconducting transition temperature,
$T_{c}$, for example in $\mathrm{ErRh}_{4}\mathrm{B}_{4}$. However, even
in this case, there is a narrow, but rich regime of the microscopic coexistence
of two quantum phases \cite{Ishikawa1977,Crabtree1982,Bulaevskii1982,Fischer1990,Prozorov2009f}.
Itinerant ferromagnetism may also coexist with superconductivity and
such materials exhibit some very unusual properties \cite{UFMSc2019}.
Most of the studied ferromagnetic superconductors are singular compositions,
which somewhat limits the possibility to study the trends and variations
of properties in the continuous phase space, such as temperature vs.
doping, $T\left(x\right)$ \cite{Fischer1990,Flouquet2002,Sinha1989}.
In contrast, there are many antiferromagnetic (AFM) superconductors where
the regime of coexistence is easier to realize. Superconductivity
develops on an AFM background as long as the internal magnetic field
modulation occurs at distances much shorter than the superconductor's
coherence length, $\xi$, which is often realized in real materials.
In turn, antiferromagnetism is largely unaffected by superconductivity
because screening of the magnetic field is effective on the length scale
of London penetration depth, $\lambda_{L}$ \cite{Buzdin1984,Buzdin1986}. However, if the FM state is formed via the RKKY exchange interaction, superconducting pairing of conduction electrons may also affect the strength of the ferromagnetic exchange. 
In general, some form of spin arrangement with a net ferromagnetic
component can be realized in a broad range of compositions in several
families of magnetic superconductors, such as borocarbides \cite{Cava1994,Budko2006,Prozorov2009e}
and more recently in some iron-based superconductors (IBS), where a
decade of intense studies have clearly shown that magnetism plays
an important, if not pivotal, role in their physics \cite{canfield2010,johnston2010,paglione2010,basov2011,hirschfeld2011,prozorov2011,Zapf2013,chubukov2015}.
In the majority of IBS magnetism arises from the iron sublattice with
spins aligned in the Fe-As plane where superconducting condensate
mostly resides. However, in a few IBS compounds, there is an additional
magnetism coming from, for example, europium as part of their formula
\cite{Zapf2013}. In $\textrm{EuFe}_{2}\textrm{As}_{2}$, Eu$^{2+}$
ions ($7\mu_{B}$ full local magnetic moment) order in an A-type antiferromagnet
below 19~K while the iron sublattice develops a spin-density-wave
(SDW) below 190~K \cite{S.Nandi2014,S.Nandi2014a}. The effect of
$\textrm{Eu}^{2+}$ magnetism is so large that can even be used to
detwin the material by applying an in-plane magnetic field \cite{Ding2020}.
Thanks to a possibility of a continuous doping of the parent compound,
superconductivity can be induced in some range of compositions
by isovalent substitution of phosphorus for arsenic. With increasing
$x$ in $\textrm{EuFe}_{2}(\textrm{As}_{1-x}\textrm{P}_{x})_{2}$,
the Eu$^{2+}$ spins become canted out of the $ab-$plane producing
a net ferromagnetic component along the $c-$axis. In our crystals
with $x=0.23$, in zero applied magnetic field, superconducting transition
occurs upon cooling at $T_{c}\left(H=0\right)\approx24\:\text{K}$,
followed by the magnetic transition of europium sublattice at $T_{m}\approx$
18 K \cite{Zapf2013}. Although rare, this is not a singular FM/SC
composition in this IBS family. In a related compound, $\text{RbEuFe}_{4}\text{As}_{4}$,
ferromagnetism develops at $T_{m}\approx15\:\text{K}$ in a superconducting
background with $T_{c}\left(H=0\right)\approx36.5\:\text{K}$ \cite{YiLiu2016,Smylie2019b}. It is worth noting that while most
theories address the coexistence of magnetism and superconductivity
in IBS with respect to the iron ions \cite{basov2011,chubukov2012,chubukov2015},
only few specifically target magnetism coming from other ions, such
as $\textrm{Eu}^{2+}$ \cite{Maiwald_PhysRevX_2018,Devizorova_2019,koshelev2020}. 

When studying complex non-stoichiometric materials it is important
to be able to fix the composition and examine the evolution of field
and temperature dependencies when some other non-thermal control parameter
is varied. One obvious example of such parameter is pressure, which
has been used intensely for this purpose. Another is a controlled disorder
that provides an important insight into magnetism and superconductivity
\cite{LYNTON1957165,Anderson1959a,AG1960,MarkowitzKadanoff1963,HirschfeldGoldenfeld,GolubovMazin1997,SunMaki1995,Openov1998,EfremovPRB2011,Kogan13pairbreaking,PdTe2eirr2020,Hc2pairbr2013}, in particular in IBS \cite{Onari2010,Gordon2010a,EfremovPRB2011,prozorov2011,TcEnhancement2012,Kogan13pairbreaking,Wang2013TcRho,Ghigo2018PRL,Torsello2019PRAppl}.
Additional scattering can be induced by various means ranging from
chemical substitution \cite{Torsello2019PRB2} to particle irradiation \cite{Torsello2020sust}. Controlled
disorder has been used to study superconductors since the times of
the famous Anderson theorem \cite{Anderson1959a} and Abrikosov-Gor'kov
theory \cite{AG1960}, where the main attention was paid
to the variation of the superconducting order parameter, hence the
experimentally accessible transition temperature, $T_{c}$ \cite{PdTe2eirr2020,Sinha1989,Maple1995,Matthias1958,D_Finnemore_1965,D.Finnemore_1967,Torsello2021SciRep}. More recently, the response to the variation of the scattering rate was studied for other properties, such as superfluid density and thermal conductivity, which are directly linked to the
superconducting order parameter structure \cite{HirschfeldGoldenfeld,Gordon2010a,EfremovPRB2011,Kogan13pairbreaking,Wang2013TcRho,Cho_2018,Torsello2019EPJST,Torsello2019JOSC}.
Due to relative rarity of magnetic superconductors, there is limited
experimental information on the effects of disorder simultaneously
on the superconductivity and magnetism. While we are not aware of
such studies in magnetic borocarbides, in IBS the effect of disorder
on magnetism and superconductivity was studied in several works \cite{Demirdi2013, Mizukami2014, Mizukami2014a, Sun2017}, but none of them on the ferromagnetic Eu-based 122 compounds, except for a
recent study of the effect of proton irradiation on the subject compound
by some of the authors \cite{Ghigo_2020}. 

\begin{figure}[tb]
\centering 
\includegraphics[width=9cm]{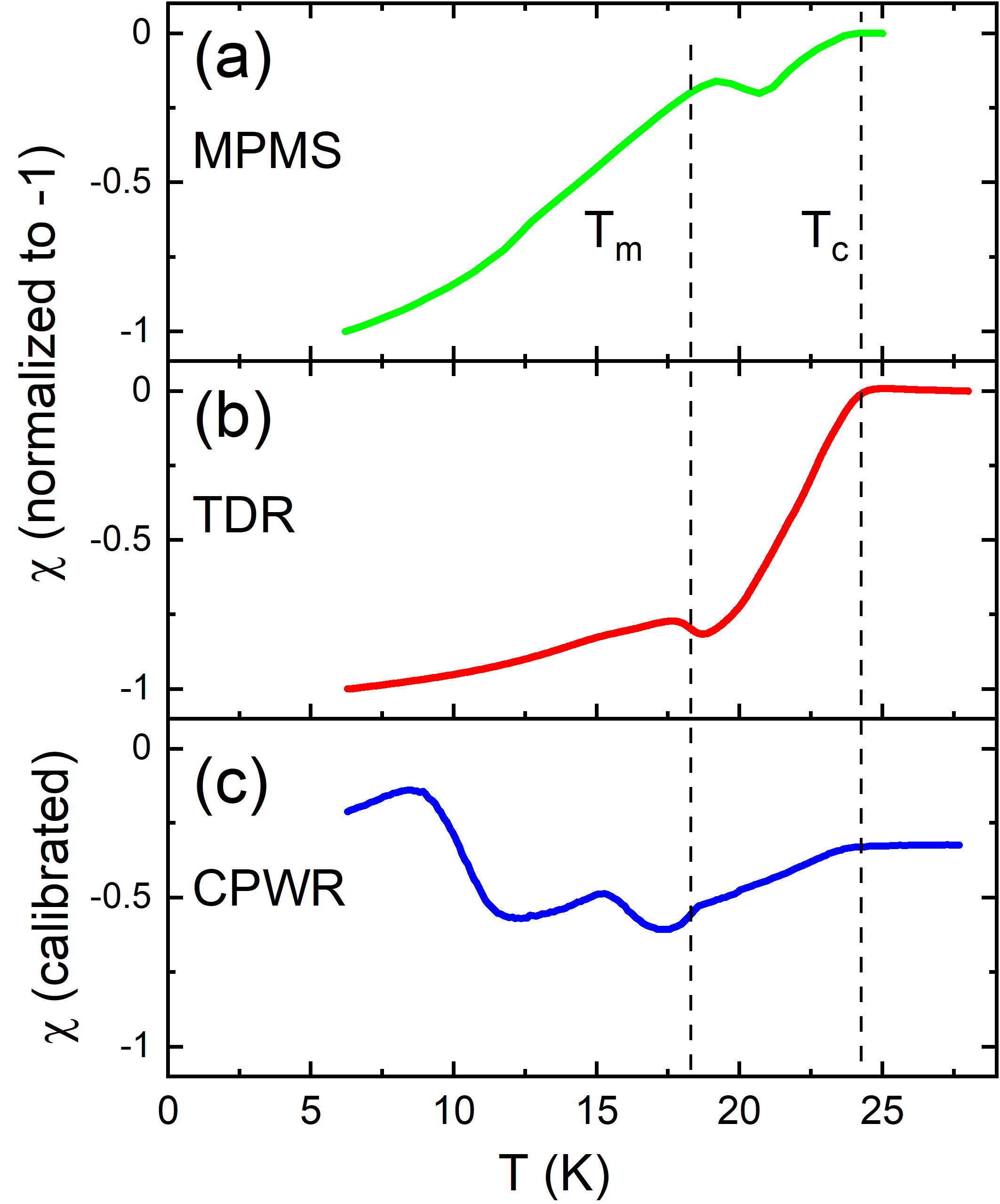} \caption{Temperature-dependent DC and real part of AC magnetic susceptibility of pristine EuFe$_{2}$(As$_{0.77}$P$_{0.23}$)$_{2}$
single crystals measured at very different frequencies: (a)  SQUID magnetometer (MPMS, Quantum Design, DC regime, $H_{DC}=5\:\mathrm{Oe}$);
(b) tunnel-diode resonator~(TDR, 14 MHz, $H_{AC}=20\:\mathrm{mOe}$),
and (c) coplanar waveguide resonator~(CPWR, 8 GHz, $H_{AC}\approx1\:\mathrm{Oe}$).
Magnetic susceptibility is normalized to $-1$ at low temperatures
for (a) and (b), but is shown in absolute values in a calibrated experiment
in panel (c).}
\label{fig1}
\end{figure}

In the present work, we study the effects of electron irradiation on single crystals of $\textrm{EuFe}_{2}(\textrm{As}_{1-x}\textrm{P}_{x})_{2}$
and compare with the proton irradiation performed on similar samples. We find that in this particular ferromagnetic superconductor, magnetic and superconducting subsystems coexist almost independently. Superconductivity interacts with the internal magnetic field produced by the Eu$^{2+}$
sublattice and the ferromagnetism is barely screened by the superconducting phase.
The non-trivial interaction is revealed when the artificial point-like disorder enhances
both potential and spin-flip scattering channels affecting $T_{c}$
at a much greater rate compared to the nonmagnetic IBS. Remarkably, controlled disorder combined with almost reversible annealing allows examining the properties of both phases in the regimes of $T_{c}>T_{m}$ and $T_{c}<T_{m}$
in a single composition.
\section{Results}

Figure \ref{fig1} shows the temperature-dependent dynamic susceptibility
of pristine $\textrm{EuFe}_{2}(\textrm{As}_{0.77}\textrm{P}_{0.23})_{2}$
single crystals measured using experimental techniques with very different
time windows. Panel (a) shows the DC results obtained by using Quantum
Design magnetic property measurement system (MPMS) at $H_{DC}=5\:\mathrm{Oe}$,
panel (b) shows 14 MHz tunnel-diode resonator (TDR) data with excitation
AC field of $H_{AC}=20\:\mathrm{mOe}$, and panel (c) shows 8 GHz
data at $H_{AC}\approx1\:\mathrm{Oe}$ obtained by coplanar waveguide resonator
(CPWR) technique.

For comparison, the data in panels (a) and (b) were normalized to
extrapolate to $\chi=-1$ at the lowest T, whereas panel (c) shows
the calibrated data. All three susceptibility curves clearly show
superconducting transition near $T_{c}\approx24\:\mathrm{K}$ and
ferromagnetic transition at $T_m\approx$18~K. The microwave-frequency CPWR
data show extra features and a detailed analysis of the measurements
is given elsewhere \cite{Ghigo_2019,Ghigo_2020}, while we are interested
in a comparison of the transition temperatures. Below $T_{c}$ diamagnetic
susceptibility is rather broad compared to much sharper transitions
of nonmagnetic superconductors. This can be attributed to a substantial
pairbreaking coming from the large-moment paramagnetic background.
In the simple picture, if ~$\mu\left(T\right)$ is the normal state
magnetic permeability, then the measured magnetic susceptibility is
renormalized as \cite{Cooper_1996,Prozorov2000i}:

\begin{equation}
\left(1-N\right)\chi\left(T\right)=\frac{\sqrt{\mu\left(T\right)}\lambda_{L}\left(T\right)}{R}\tanh\frac{\sqrt{\mu\left(T\right)}R}{\lambda_{L}\left(T\right)}-1
\end{equation}

\noindent where $N$ is the effective demagnetizing factor, $R$ is the effective dimension and $\lambda_{L}\left(T\right)$ is the London penetration depth without magnetism present. Below $T_{c}$
the $\sqrt{\mu\left(T\right)}\lambda_{L}\left(T\right)$ term dominates
the behavior with two competing trends. Taking the simplest functional
forms, in the interval $T_{m}<T<T_{c}$,

\begin{equation}
\sqrt{\mu\left(T\right)}\lambda_{L}\left(T\right)\sim\left(\left(T-T_{m}\right)\left(T_{c}-T\right)\right)^{-1/2}
\end{equation}

\noindent which is, indeed, a non-monotonic function of temperature in this
interval which is seen in all three measurements shown in Fig.~\ref{fig1}.
Below $T_{m}$ the magnetic susceptibility decreases and the overall
signal tends to decrease again. Of course, the physics around ferromagnetic
transition is significantly affected by the proliferation of spontaneous
(vortex - antivortex) phases as was determined in the comprehensive microscopic study
\cite{Stolyarov2018}. This scenario has been further explored in
\cite{Ghigo_2019,Devizorova2019} Similarly, effects of spontaneous
vortex phase was investigated both experimentally and theoretically
in already mentioned 1144 sibling compound, $\text{RbEuFe}_{4}\text{As}_{4}$
\cite{YiLiu2016,Koshelev2019a,Vlasko2019,Smylie2019b}.

\begin{figure}[tb]
\centering \includegraphics[width=9cm]{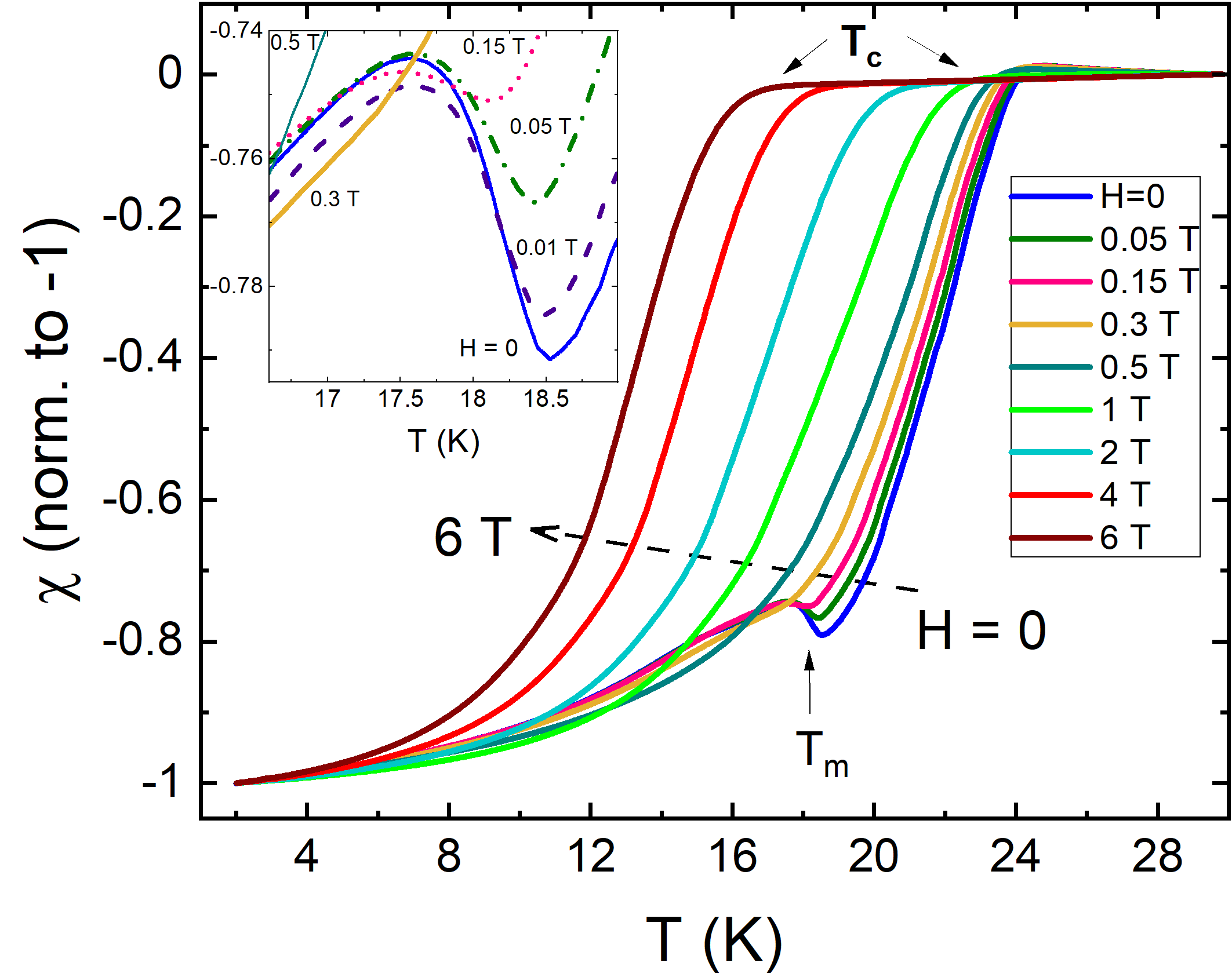} \caption{Normalized magnetic susceptibility of pristine EuFe$_{2}$(As$_{1-0.23}$P$_{0.23}$)$_{2}$
single crystal from TDR measurements at different DC magnetic fields
applied along the $c-$axis. The inset zooms at the rapid suppression
of the peak near the ferromagnetic transition.}
\label{fig2}
\end{figure}

Figure \ref{fig2} shows TDR measurements of temperature dependent
magnetic susceptibility of pristine $\textrm{EuFe}_{2}(\textrm{As}_{0.77}\textrm{P}_{0.23})_{2}$
at different magnetic fields applied along the $c-$axis. As expected,
$T_{c}\left(H\right)$ decreases with the increasing magnetic field while $T_{m}$
remains practically unchanged. However, as shown in the inset in Fig.~\ref{fig2}, at the same time the height of the peak near $T_{m}$ decreases and disappears completely above 0.2~T.
This is a characteristic behavior associated with a local moment ferromagnetism
as shown previously using TDR technique \cite{vannette2008}.
 
\subsection{Electron irradiation}

We now turn to the effects of the artificial disorder induced by the 2.5
MeV electron irradiation. Details of the experiment are described
in section \ref{sec:Materials-and-Methods}. The irradiation dose is measured
during the experiment as a total charge flux passed through the sample
and can be expressed in convenient practical units of coulomb per cm$^2$ to represent the irradiation
dose, $1\: \textrm{C/cm}^2 = 6.24 \times 10^{18}\: \textrm{electrons}/\textrm{cm}^{2}$.

\begin{figure}[tb]
\centering \includegraphics[width=9cm]{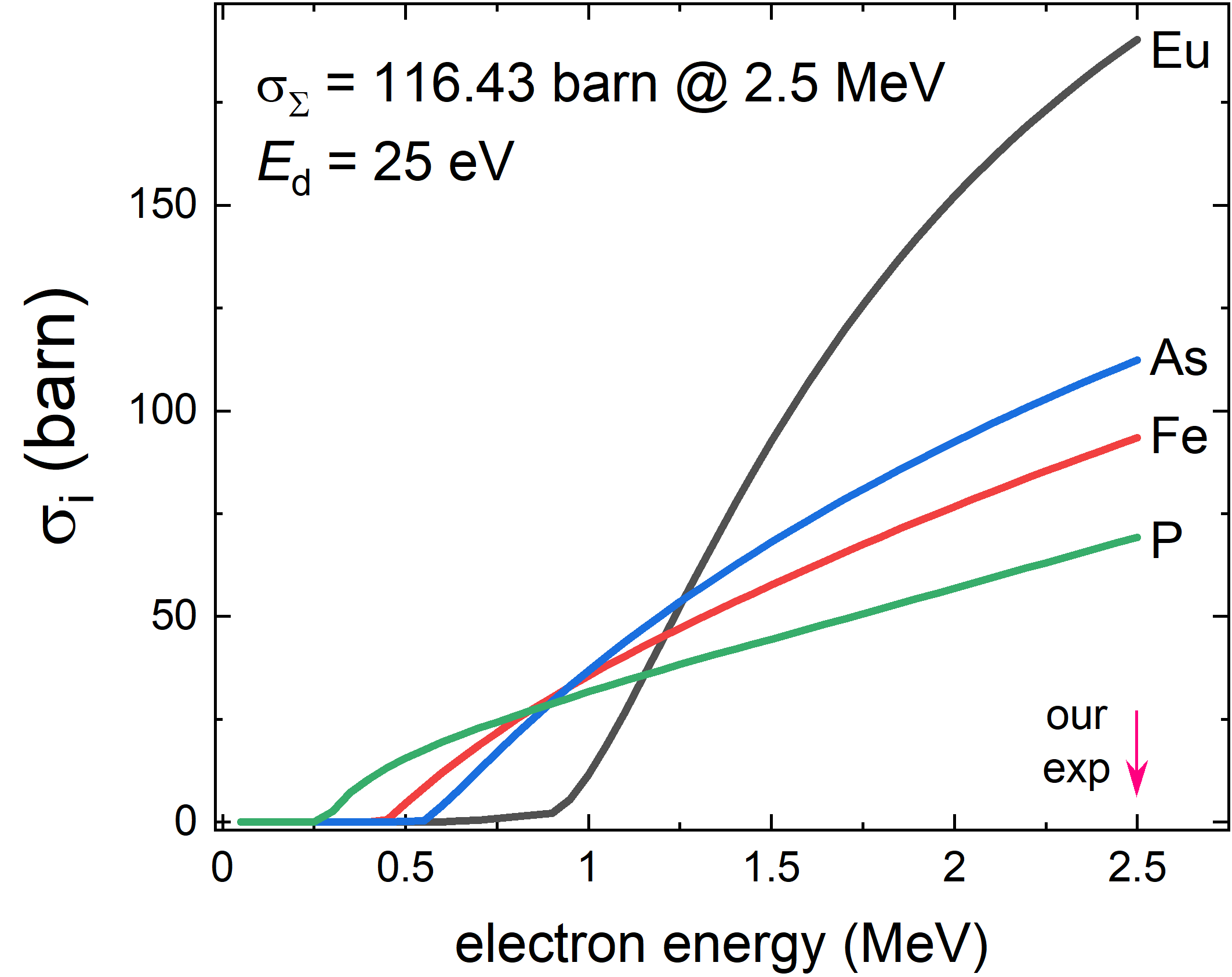} \caption{Defects creation cross-sections for different ions in $\textrm{EuFe}_{2}(\textrm{As}_{0.77}\textrm{P}_{0.23})_{2}$
as function of electron energy assuming the displacement energy threshold,
$E_d=25\:\mathrm{eV}$. 
At 2.5 MeV, the partial cross-sections are
P: 69.2 barn, Fe: 93.5 barn, As: 112.4 barn and Eu: 190.3 barn. The total
cross-section of defects creation is estimated as 116.4 barn, which
leads to the estimate of the $7.3 \times 10^{-4}$ displacements-per-atom
(dpa) per 1 C/cm$^2$ of the irradiation.}
\label{fig3}
\end{figure}

Figure \ref{fig3} shows partial cross-sections of the defects creation
calculated using SECTE software, developed at Ecole Polytechnique
(Palaiseau, France) specifically for electron irradiation. 

Of course, the greatest uncertainty is the displacement threshold
energy barrier, $E_{d}$, which varies between 10 and 50 eV for different
ions and compounds \cite{Damask1963,Konobeyev2017,Torsello2018PRMat}. In this work its precise value is not important since
we only need the order of magnitude estimate. We used a typical
value of 25 eV commonly assumed for cross-section calculations for both electron and proton irradiations \cite{Damask1963}.
This gives around 0.07 at.\% dpa (displacements-per-atom) per $\mathrm{1\:C/cm}{}^{2}$
of electron irradiation or about 7 defects-creating collisions per 1000
unit cells (10 atoms in a \textit{Z}=2 unit cell) and about twice that value for protons. Therefore the density of the defects is small and they do not alter chemical composition and
do not ``dope'' the system, which was proven by Hall effect measurements
in another 122 compound, BaK122 \cite{Prozorov2019}. Examination
of Fig. \ref{fig3} shows that irradiation at our energy of 2.5 MeV
produces mostly defects on the Eu sites, whereas lower energy, say
1 MeV, would produce the least defects on the Eu sites. Such energy-tuneable
irradiation is possible and would lead to ion-specific study of the
effects of disorder.

\begin{figure}[tb]
\centering \includegraphics[width=9cm]{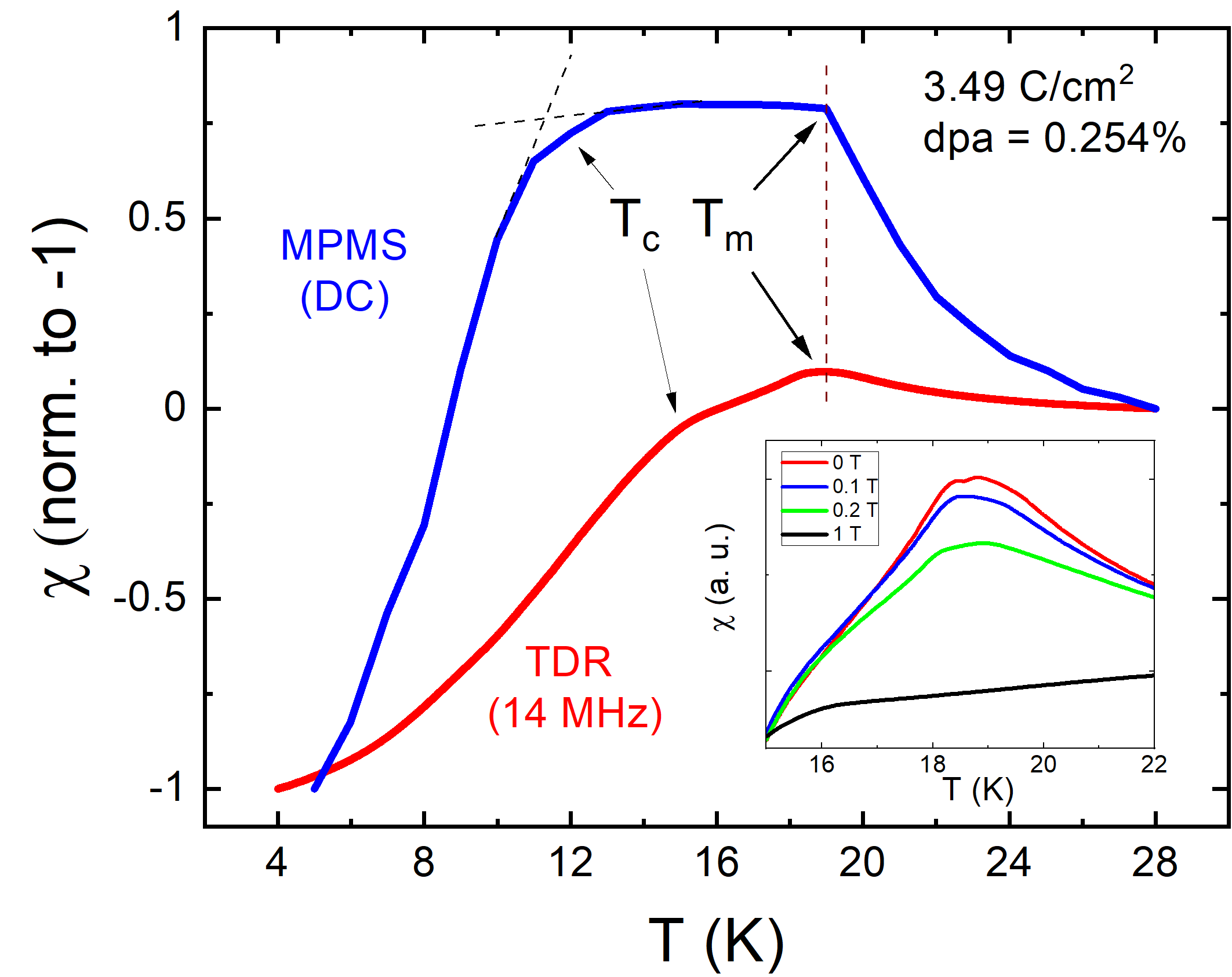} \caption{Magnetic susceptibility of $\textrm{EuFe}_{2}(\textrm{As}_{0.77}\textrm{P}_{0.23})_{2}$ single crystal after 2.5 MeV electron irradiation with a dose of 3.49 C/cm$^{2}$ (0.254
at.\% dpa) measured in a DC regime using MPMS (top blue curve) and at
14 MHz using TDR (red curve). The inset shows the evolution of the
TDR peak near the ferromagnetic transition for different applied magnetic
fields.}
\label{fig4}
\end{figure}

It is important to note that we studied physically the same crystals before
and after the irradiation, so the observed changes are the results of the added
disorder. Figure \ref{fig4} shows the temperature dependent susceptibility
of electron irradiated sample with the dose of 3.49 C/cm$^{2}$
measured using MPMS~(blue curve) and TDR~(red curve). Both measurements
clearly show a very significant $T_{c}$ suppression, but only a modest increase of 
$T_{m}$. This leads to an outstanding result that the irradiation has
driven the superconducting transition from well above $T_m$ to
well below. Therefore, we have a unique situation that both regimes could be
studied in the same sample. One of the important properties is the
transition temperature itself as the function of disorder. While in the
regime of $T_{c}>T_{m}$ both measurements gave similar $T_{c}$,
see Fig.~\ref{fig1}, we see very a different $T_{c}$ in the irradiated
sample measured by the two techniques when $T_{c}<T_{m}$. Clearly,
the difference is due to the dynamic nature of the superconducting
transition. Now the superconductivity develops on a ferromagnetic background,
hence in the presence of a finite internal magnetic field and, therefore, the nature of the transition is reminiscent of the magnetic irreversibility temperature, which is known to be very
frequency-dependent \cite{Prozorov1995} in materials with large
magnetic relaxation, such as high$-T_{c}$ cuprates \cite{Yeshurun1996,Blatter1994}
and iron pnictides \cite{VortexBaK122PRB2012,Prozorov2009d,Prozorov2008b}.

\begin{figure}[tb]
\centering \includegraphics[width=9cm]{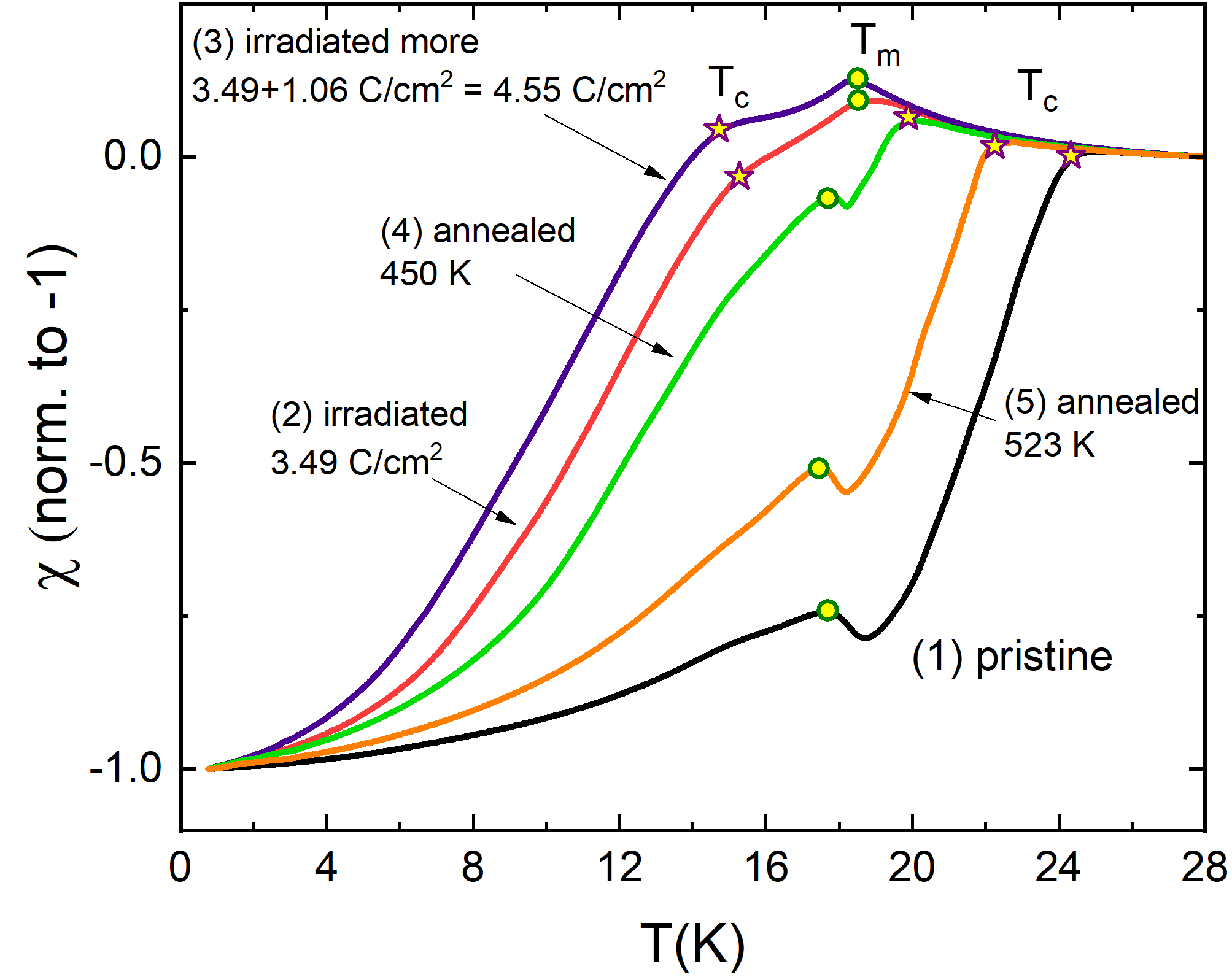} \caption{Electron irradiation and subsequent annealing studies of the same $\textrm{EuFe}_{2}(\textrm{As}_{0.77}\textrm{P}_{0.23})_{2}$
sample as shown in Fig.\ref{fig4}. Curve (1): pristine state; (2):
after 3.49 C/cm$^{2}$ irradiation; (3) 4.55 C/cm$^{2}$ total dose
where 1.06 C/cm$^{2}$ was added after step (2); (4) after annealing
at 450 K, and (5) after the second annealing at 523 K. Stars mark the superconducting transition and the circles mark the ferromagnetic transition.}
\label{fig5}
\end{figure}

In the previous studies, we found that defects introduced by the electron irradiation
can be annealed leading to the recovery towards the pristine state, sometimes almost completely \cite{Serafim_2016, CaK1144eirr2018}. Figure \ref{fig5} shows the evolution of the dynamic magnetic susceptibility measured using TDR first after
two subsequent irradiation runs and then after two steps of annealing. Curve
(1) shows the pristine state; (2): after 3.49 C/cm$^{2}$ irradiation;
(3) at 4.55 C/cm$^{2}$ total dose where 1.06 C/cm$^{2}$ was added
after the preceding step; (4) after annealing at 450 K, and (5) after
the second annealing at 523 K. The annealing was done in argon atmosphere for several hours and then cooling overnight before opening the chamber. We observe a remarkable practically reversible transformation from the initial state with $T_{c}>T_{m}$ to the state
with $T_{c}<T_{m}$ and back to $T_{c}>T_{m}$ again. The stars mark the apparent superconducting transition and the circles mark the ferromagnetic transition. Therefore, the superconducting state can be switched off by the electron irradiation and recovered by the annealing, leaving magnetism practically intact, thanks to the local nature of Eu moments. It is quite different in the case of itinerant magnetism of iron where
the magnetic transition is suppressed at the same large rate as the superconducting transition \cite{BaRu122PRX2014}. Here, $T_m$ slightly increases by less than a degree when magnetism sets in in the normal metal. This shows that superconductivity weakens (screens) the exchange interaction suggesting, although indirectly, that two phases coexist microscopically.

\begin{figure}[tb]
\centering \includegraphics[width=9cm]{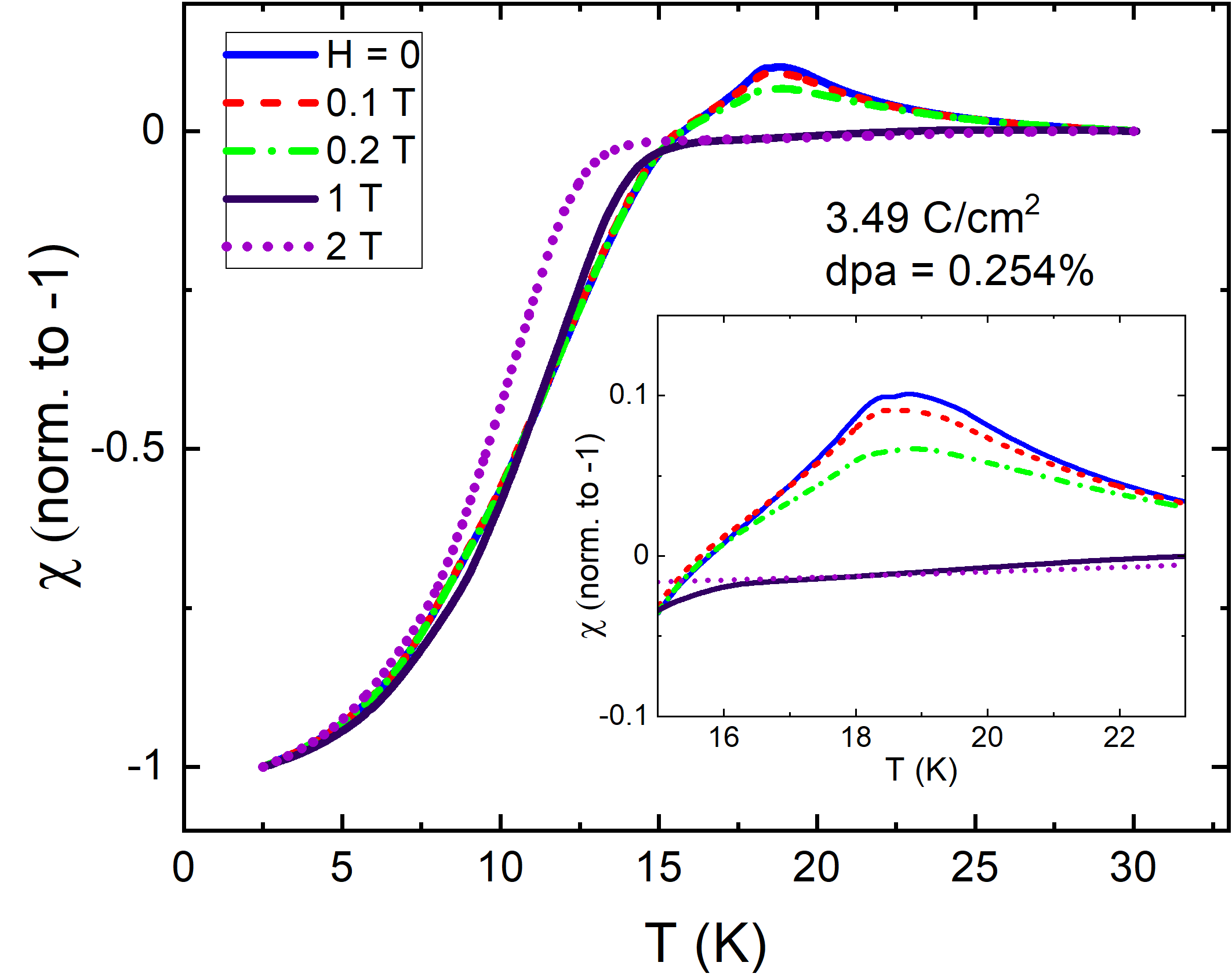} \caption{The magnetic-field dependence of TDR dynamic magnetic susceptibility
in a 3.49 C/cm$^{2}$ electron irradiated sample. The indicated magnetic
fields were applied parallel to c-axis. The inset zooms near the transition
region.}
\label{fig6}
\end{figure}

When studying superconductors, it is often needed to reveal the behavior of the normal state ``behind'' the superconducting response. For example, to estimate the phonon contribution to the specific heat. A common
recipe is to apply strong enough magnetic field and suppress superconductivity. 
However, in ferromagnetic superconductors, with relatively high $T_m$ and $T_c$, the specific heat jump at the superconducting transition temperature, is hardly detected/resolved, since the magnetic contribution to the specific heat can be large \cite{S.Nandi2014,PRL2009}. Therefore, our results provide an alternative method to reveal the normal state and, if needed, recover back the superconducting state by annealing. Furthermore, this way of $T_c$ suppression by the irradiation without altering chemical composition can be applied for quantitative specific heat studies of other ferromagnetic superconductors even with with $T_c \leq T_m$, in order to deduce magnetic and superconducting volume fractions by moving $T_c$ further down to show that both FM and SC phases are bulk in nature (or not). This also allows studying the influence of moderate magnetic fields on the FM transition that is linked to the character
of magnetism \cite{vannette2008}. By suppressing the superconducting
state by electron irradiation we reveal the local nature of ferromagnetism
in $\textrm{EuFe}_{2}(\textrm{As}_{0.77}\textrm{P}_{0.23})_{2}$. This follows from the behavior of the peak in dynamic susceptibility in the normal state. Figure \ref{fig6} shows the evolution of a ferromagnetic peak with
the applied DC magnetic field along c-axis. Upon cooling from above
$T_{m}$, TDR measurements in local-moment systems exhibit a sharp
peak in zero field. When a small magnetic field is applied, it reduces
the amplitude of the peak as shown in the inset in, Fig.\ref{fig6}.
In case of itinerant ferromagnetism there is broad maximum rapidly
smearing and shifting to the lower temperatures \cite{vannette2008}.

\begin{figure}[tb]
\centering \includegraphics[width=9cm]{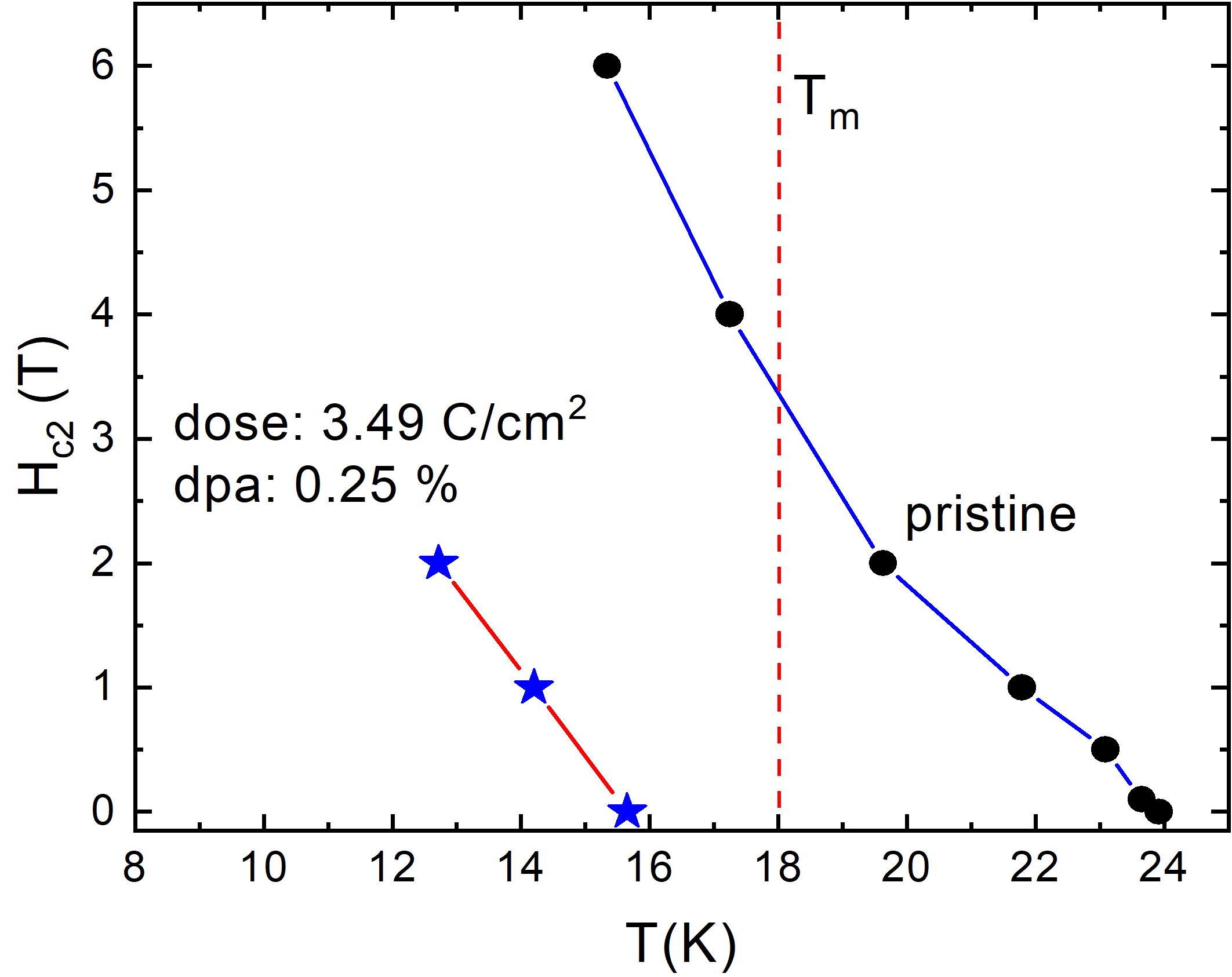} \caption{Upper critical field, $H_{c2}(T)$, of pristine (black circles) and 3.49 C/cm$^{2}$ electron irradiated (blue stars) $\textrm{EuFe}_{2}(\textrm{As}_{0.77}\textrm{P}_{0.23})_{2}$ with the magnetic field applied along the $c-$axis. Note that the slope, $dH_{c2}(T)/dT$ near $T_{c}$ in irradiated state is larger than the slope in the pristine state.}
\label{fig7}
\end{figure}

Finally, we compare the upper critical field, $H_{c2}(T)$, in pristine state (black filled circles in Fig.~\ref{fig7}) and after 3.49 C/cm$^{2}$ electron irradiated state (blue stars) of the same sample. The data for the pristine samples are close to the values reported for polycrystalline sample \cite{PRL2009}. While it is expected that $H_{c2}(T)$ may
have a step-like feature at $T_{m},$ we did not have an opportunity to study the 
$H_{c2}(T)$ line in great detail and it is impossible to draw any
conclusions from our data. Yet, the curve shows an unusual positive
curvature enetring the region of $T \leq T_{m}$, which is not expected in standard
models  \cite{Kurita_2011}. We can speculate that the magnetic pair-breaking scattering is reduced in the long-range ordered phase, because it requires spin-flip of
the scatterer. This will cause an increase of $H_{c2}(0)$ \cite{Hc2pairbr2013}.
\noindent Furthermore, the slope, $dH_{c2}(T)/dT$ near $T_{c}$ is proportional
to $T_{c}$ multiplied by a function of potential and pair-breaking
scattering \cite{Hc2pairbr2013}. That function increases with the
increase of the potential scattering and decreases with the increase of
the pair-breaking one. According to the Anderson theorem, potential scattering
does not change $T_{c}$ whereas pair-breaking scattering decreases
$T_{c}$. Note that in sign-changing $s_{\pm}$ order parameter, the
interband potential (non spin-flip) scattering is also pair-breaking, while the inband potential scattering is not, provided that each band has no nodes or significant anisotropy. \cite{Kogan2016}. Figure \ref{fig7} shows that the slope at lower $T_{c}$ is actually larger than that in the larger
$T_{c}$ pristine state indicating that the pair-breaking scattering increases $H_{c2}$ faster
than it suppresses $T_{c}$ adding to the conclusion that electron
irradiation produces a substantial amount of the additional pair-breaking
scattering.

\subsection{Phase diagram and comparisons with other compounds and irradiation}

Indeed, the response to any perturbation, irradiation included, should
be gauged against the results obtained with other types of materials
and irradiations. Here we compare the results with CPWR measurements
of proton-irradiated samples. Protons also introduce largely point
like defects and, in addition, nanometric clusters, which slightly reduce the efficiency of the overall produced defects acting as scattering centers. Detailed account of the effects of disorder by doping and proton irradiation in $\textrm{EuFe}_{2}(\textrm{As}_{0.77}\textrm{P}_{0.23})_{2}$ is given elsewhere \cite{Ghigo_2020}. 

\begin{figure}[tb]
\centering \includegraphics[width=9cm]{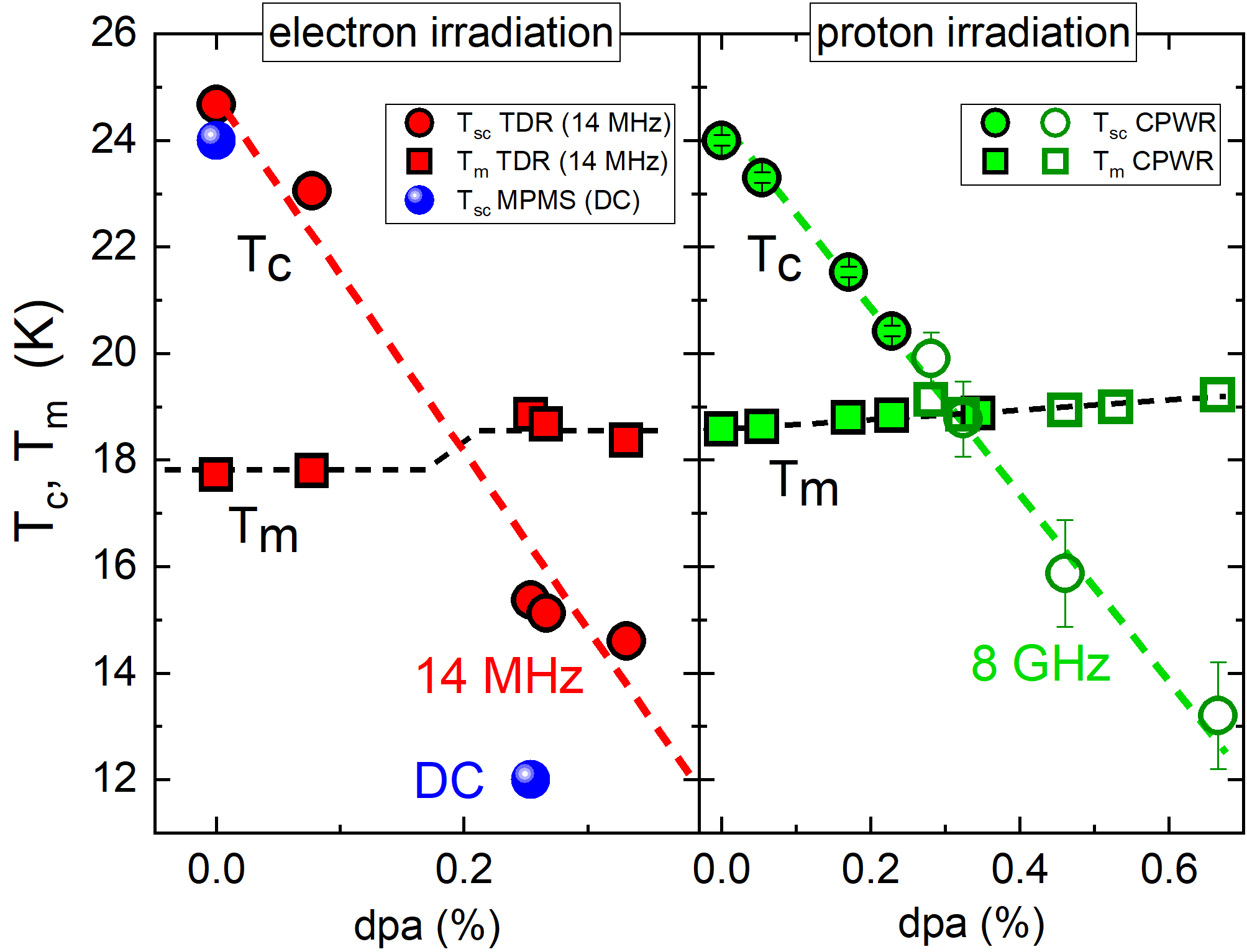} 
\caption{Transition temperatures vs. the estimated atomic concentration of defects in $\textrm{EuFe}_{2}(\textrm{As}_{1-x}\textrm{P}_{x})_{2}$ single crystals obtained by different techniques and different doses of irradiation. (a) electron irradiation in $x=0.23$ crystal, measurements using (red)  tunnel-diode resonator (TDR) and (blue) DC magnetometry; (b) proton irradiation in $x=0.23$ (hollow symbols) and $x=0.20$ (full symbols) crystals, measurements using coplanar waveguide resoantor (CPWR). The way the effective disorder has been estimated in this case was discussed in \cite{Ghigo_2020}.}
\label{fig8}
\end{figure}

\begin{figure}[tb]
\centering \includegraphics[width=9cm]{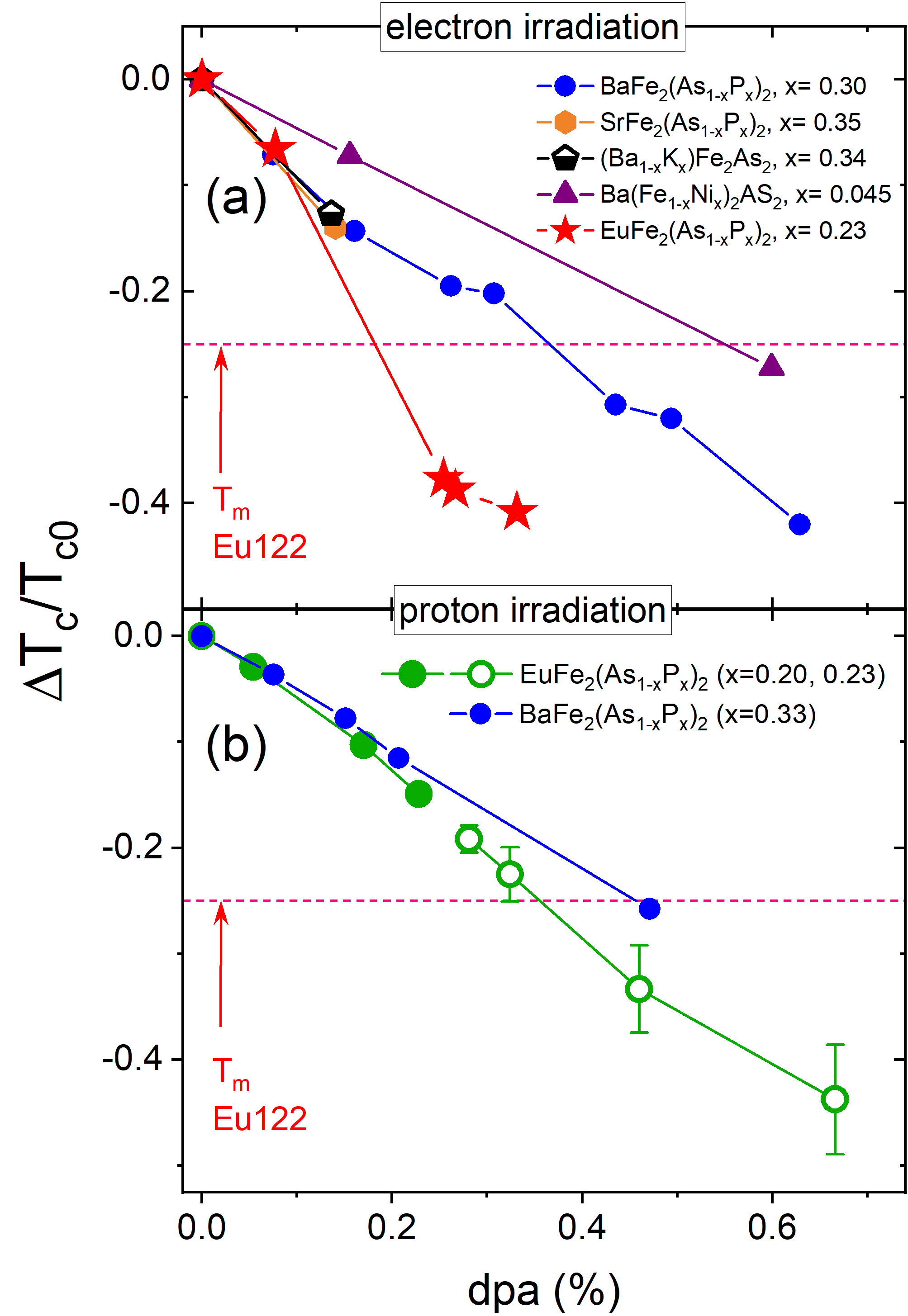} \caption{The normalized rate of $T_{c}$ suppression by electron irradiation (panel (a)) and proton irradiation (panel (b)). Electron irradiation results from this work are compared to the data in other compounds, summarized in Ref.~\cite{Cho_2018}. The data for the important for direct comparison $\textrm{BaFe}_{2}(\textrm{As}_{1-x}\textrm{P}_{x})_{2}$ are from Ref.~\cite{Mizukami2014}. The dpa percentages were calculated for each of the listed compounds using
SECTE calculations similar to Fig.\ref{fig3}. For CPWR data, hollow symbols were obtained from a crystal with additional doping-induced disorder (x=0.23). The way the effective disorder has been estimated and its data merged to that of the x=0.20 was discussed in \cite{Ghigo_2020}.}
\label{fig9}
\end{figure}

Figure \ref{fig8} shows the superconducting and ferromagnetic transition
temperatures versus the estimated atomic percentage of induced defects. The displacements-per-atom (dpa) values are based on the calculated cross-sections using SECTE for electron
irradiation and SRIM for proton irradiation \cite{Ziegler2010}. We therefore present the data on two panels, since the dpa values are obtained using different calculations.
Regardless of the type of irradiation or experimental time window, the magnetic transition remains robust and stays around
$T_m\approx$~18~K only slightly increasing in the normal state. The superconducting transition, while suppressed significantly, decreases at a similar rate for different experimental methods operating at very different frequencies as long as $T_{c}>T_{m}$. However, the agreement breaks after the transition temperatures swap places, $T_{c}<T_{m}$. There are two major contributions at play here. One is the time window of the measurement,  which leads to lower $T_{c}$ for smaller frequencies (longer relaxation).
The second parameter is the nature of the defects. Considering that
electrons and protons have not only very different masses, but also
opposite signs of their charge. Unfortunately, we do not have ion-type
resolved particle cross-sections for the proton irradiation,
but it is quite possible that protons knock out particular ions at
rates different from electrons. Here we can conclude that the background
magnetism affects superconductivity in a way consistent with the conclusions
of the previous studies - the superconducting phase develops with the
magnetic field present, which immediately triggers significant time-dependencies
of all measurable parameters.

Figure \ref{fig9} compares the normalized rate of the $T_{c}$ suppression,
$\Delta T_{c}/T_{c0}$, plotted versus the estimated density of the induced
defects, which was calculated for each of the listed compounds. Similar to \ref{fig9}, two panels show the results for electrons and protons, respectively. Panel (a) summarizes the results of electron irradiation, while panel (b) shows proton irradiation. We stress that the estimated dpa values are estimated using two different approaches and, also, do not take into account recombination upon warming and possible clusterization and agglomeration into larger non-point like groups. Further controlled studies on similar samples are needed to compare electron and proton irradiation in a quantitative way.
Here we see that the suppression rate for electron-irradiated $\textrm{EuFe}_{2}(\textrm{As}_{0.77}\textrm{P}_{0.23})_{2}$ is higher
compared to others, non-magnetic, compounds of the IBS. Most likely
this is because of the formation of magnetic scattering centers on
Eu sites, in addition to the non magnetic channel formed by all defects.
Considering partial cross-sections shown in Fig.~\ref{fig3} this scenario
is quite plausible. In the case of $\textrm{BaFe}_2(\textrm{As}_{1-x}\textrm{P}_{x})_{2}$ irradiated by 3 MeV protons, we also found that the suppression rate of $T_c$ is
larger compared with similar compounds. We attribute this enhanced suppression of $T_c$ to the generation of Fe$_{2}$P, which is one of possible magnetic compounds that can also generate magnetic scattering \cite{Park_2018}.

\section{Conclusions}

In summary, we used controlled disorder produced by electron and proton
irradiations to induce two states in the same sample: (1) $T_{c}>T_{m}$
and (2) $T_{c}<T_{m}$. The ferromagnetic transition, $T_{m}$, is weakly affected
by the irradiation whereas the superconducting transition, $T_{c}$, is
rapidly suppressed. We therefore had a unique opportunity to study
the same ferromagnetism in normal and superconducting background and, vice versa,
superconductivity developing in a paramagnetic or a ferromagnetic background. The ferromagnetic transition temperature increases by less than a degree in the normal state compared to when it is born out of superconducting background signaling of the microscopic coexistence and, perhaps, some weakening of the exchange interaction by the superconducting phase. 
Furthermore, we conclude that in $\textrm{EuFe}_{2}(\textrm{As}_{1-x}\textrm{P}_{x})_{2}$ local-moment ferromagnetism of Eu$^{2+}$ sublattice does not have a direct impact on superconducting pairing,
but it affects the superconducting state via the spontaneous internal
magnetic field that creates Abrikosov vortices and antivortices in the neighboring domains. 
When $T_{c}<T_{m}$, the superconducting transition becomes significantly frequency-dependent reminiscent of the irreversibility
temperature, $T_{irr}(H)$, rather than the true zero-field transition $T_{c}(H=0)$. 
It is also possible that the annihilation of vortex-antivortex pairs at the domain boundaries assisted by an AC field at $T_{c}<T_{m}$ can further enhance the frequency dependence due to the dynamic response of shaking-depinned vortex-antivortex lattice, which is different from that of a conventional mixed vortex state. Another effect of Eu$^{2+}$ sublattice is to provide the effective pair-breaking ``magnetic''
defects upon particle irradiation. This leads to even faster $T_{c}$
suppression by disorder than in non-magnetic 122 compounds. This also
means that the pairing state of $\textrm{EuFe}_{2}(\textrm{As}_{1-x}\textrm{P}_{x})_{2}$ is most likely $s_{\pm}$ as in other IBS.
 
\section{Materials and Methods}
\label{sec:Materials-and-Methods}
\noindent \textbf{Crystal growth, samples}. Single crystals of $\textrm{EuFe}_{2}(\textrm{As}_{1-x}\textrm{P}_{x})_{2}$,
x = 0.23, were grown using self flux method from Eu powder (3N purity), FeAs and FeP precursors, mixed stoichiometrically with nominal x = 0.25 \cite{XiaofengXu2014}. The batch was grown inside stainless steel tube in nitrogen atmosphere with $T_{max} = 1350 ^{\circ} C$ (heating at a rate of 50 $^{\circ}$C/hour, keeping at $T_{max}$ for 12 hours) followed by slow cooling at 2 $^{\circ}$C/hour down to $T_{min} = 1000 ^{\circ}C$.

\noindent \textbf{Tunnel Diode Resonator(TDR)}. The real part of the radio-frequency magnetic susceptibility was measured
by using a sensitive tunnel diode resonator (TDR) \cite{Prozorov2000,Prozorov2000a,Prozorov06}.
The sample (typically $\sim0.5\times0.5\times0.1\:\mathrm{mm}^{3}$)
is mounted on a sapphire rod using a trace amount of Apiezon\textsuperscript{TM}
N-grease in a desired orientation and inserted into the inductor coil.
The coil generates a small AC excitation magnetic field, $H_{ac}\approx1-10\:\mathrm{A/m}$,
the exact value of which depends on the distance between the coil and a copper
tube in which the coil is housed for temperature stability and electromagnetic
shielding. The other end of the sapphire rod is glued into a copper
block containing a Cernox\textsuperscript{TM} thermometer and a resistive
heater.

In the experiment, the resonant frequency of the $LC$ tank circuit
with the sample inside the coil is recorded as function of temperature
or external DC magnetic field, generated by the superconducting magnet
outside the cryostat. The shift of the resonant frequency, $\Delta f=f(H,T)-f_{0}$,
from its value without the sample, $f_{0}$, is proportional to the
sample magnetic susceptibility \cite{Prozorov2000,Prozorov2021}:

\begin{equation}
\Delta f\equiv f(H,T)-f_{0}=-\frac{f_{0}V_{s}}{2V_{c}}\chi(H,T)
\end{equation}
\noindent where $\chi(H,T)=dM/dH$ is the actual magnetic susceptibility of
a given sample with volume magnetization $M=m/V_{s}$, where $m$
is total measured magnetic moment. In paramagnetic samples $\chi>1$,
then the total inductance of the sample inside the coil increases,
and resonant frequency decreases, whereas in a diamagnetic sample
the opposite is true. In a superconducting sample, the magnetic susceptibility
of a superconductor is given approximately by \cite{Prozorov2000,Prozorov2021}

\begin{equation}
\left(1-N\right)\chi(H,T)\approx\frac{\lambda}{R}\tanh\frac{R}{\lambda}-1
\end{equation}
\noindent where $N$ is the effective demagnetizing factor \cite{Demag2018}
and $R$ is the effective dimension calculated numerically for a particular
sample geometry \cite{Prozorov2000}. Considering a superconducting
sample with magnetic penetration depth $\lambda\ll R$, where $2R$
is the size of the sample in the direction of magnetic field penetration (field penetrates from two sides),
we obtain for the penetration depth:

\begin{equation}
\Delta\lambda\equiv\lambda(H,T)-\lambda(0,0)\approx R\frac{2V_{c}\left(1-N\right)}{f_{0}V_{s}}\Delta f=G\Delta f
\end{equation}
\noindent where $G$ is the calibration constant.
The main source of uncertainty in $G$ is the approximate factor
\begin{equation}
\Delta f_{0}=\frac{f_{0}V_{s}}{2V_{c}\left(1-N\right)}\label{eq:df0}
\end{equation}
\noindent which gives the change in frequency when an ideal diamagnetic sample
of the same shape and volume as the sample under study is inserted
at base (theoretically at zero) temperature into the coil. The approximate Eq.\ref{eq:df0}
is based on an idealized picture of an infinite solenoid where the sample
perturbs the magnetic flux inside. For a realistic finite coil, Eq.\ref{eq:df0}
is only a rough approximation. However, $\Delta f_{0}$ can be measured
directly by mechanically pulling the sample out of the coil at the
base temperature. Our cryostat is equipped to do just that, so we
determine $\Delta f_{0}$ directly for each sample. Then,
\begin{equation}
G=\frac{R}{\Delta f_{0}}
\end{equation}
\noindent which shows the simplified meaning of constant $G$ as the frequency
shift when magnetic field penetrates the entire sample (of size $2R$,
- from two sides, travelling distance $R$ from each side).

The measurement was conducted down to 400 mK using Janis wet Helium-3 cryostat and a DC magnetic field that can be provided by the superconducting magnet ranges up to 9 T. Further details and applications of TDR technique can be found elsewhere \cite{Prozorov2000,Prozorov2000a,Prozorov06,Prozorov2021}.\\

\noindent \textbf{Coplanar Waveguide Resoantor~(CPWR)}.The coplanar-waveguide-resonator technique allows determination of the complex permeability of small samples coupled to the resonator,  within a cavity perturbation approach\cite{Ghigo2017scirep,Torsello2019prb}. The presence of the sample coupled to the resonator induces changes in the resonance frequency and quality factor of the CPWR, that are related to the real and imaginary parts of the total AC susceptibility, respectively\cite{Ghigo_2020}:\\

\begin{align}
\Re\chi&\approx 1-\frac{2\Delta f/f_0}{\Gamma_f}\label{deltafsuf}\nonumber\\
\Im\chi&\approx \frac{\Delta\left(1/Q\right)}{\Gamma_Q}
\end{align}
where $\Delta f/f_0$ and $\Delta(1/Q)$ are the experimental shifts of the resonance frequency and of the inverse of the quality factor induced by the presence of the sample under test, and $\Gamma_f$ and $\Gamma_Q$ are geometrical factors that can be determined by a self-consistent procedure, which  takes into account also the finite size of the crystal and the consequent demagnetization effects \cite{Ghigo2017prb}. The overall real and imaginary parts of the susceptibility for ferromagnetic superconductors are given by a bulk magnetic contribution and by a screening given by the superconducting condensate.
The superconducting transition temperature $T_c$ corresponds to the onset of the diamagnetic signal (onset of an increase of the resonance frequency upon cooling), while the magnetic transition temperature $T_m$ is defined as the onset of a positive contribution to the bulk susceptibility \cite{Ghigo_2019}.\\

\noindent \textbf{Electron irradiation.} The 2.5 MeV electron irradiation was performed at the SIRIUS
Pelletron linear accelerator facility operated by the Laboratoire
des Solides Irradies (LSI) at the Ecole Polytechnique in Palaiseau,
France. At 2.5 eV  electrons are moving with relativistic speed
of 0.985$c$ and the total flux of electrons is about 2.7 $\mu$A
of electric current through a 5 mm diameter diaphragm. The acquired
irradiation dose is measured by a calibrated Faraday trap behind the
sample and is conveniently expressed in $\textrm{C}\cdot\textrm{cm}^{-2}$,
where 1 $\textrm{C}\cdot\textrm{cm}{}^{-2}=6.24\times10^{18}$ electrons/cm$^{2}$.
Electrons are particularly useful, because unlike heavier particles,
they produce well-separated point like defects, called Frenkel pairs
(vacancy+interstitial). But even with electrons, the irradiation needs
to be conducted at low temperature, in liquid hydrogen in our case
to prevent rapid clusterization of newly formed Frenkel pairs. Upon
warming up the interstitials leave the system via various sinks, such
as surfaces, defects, dislocations etc and a metastable population
of vacancies remains. Their concentration is determined by the highest
temperature reached - we re-checked the irradiated samples after a
year on the shelf at room temperature with no noticeable change. We
describe the annealing experiment in the main text. In 122 IBS, we
estimate that warming up from 22 K of irradiation run to the room
temperature, about 70\% of induced scattering centers survives as
determined from in-situ resitivity measurements \cite{BaRu122PRX2014,Prozorov2019}.\\

\acknowledgments{We thank A. Koshelev for useful discussions and the EMIR SIRIUS team,
O. Cavani, B. Boizot, V. Metayer, and J. Losco, for running the electron irradiation. Work in Ames was supported by the U.S. Department of
Energy (DOE), Office of Science, Basic Energy Sciences, Materials
Science and Engineering Division. Ames Laboratory is operated for
the U.S. DOE by Iowa State University under contract DE-AC02-07CH11358.
Work in Torino was partially supported by the Italian Ministry of
Education, University and Research (Project PRIN ‘HIBiSCUS,’ Grant No. 201785KWLE). The 2.5 MeV electron irradiation was performed at the
``SIRIUS'' facility in Ecole Polytechnique, Palaiseau, France, a
part of EMIR\&A French national network of accelerators under proposals
num. 15-5788,16-4368,18-5155. The 3.5 MeV proton irradiation was performed at the CN facility of the Legnaro National Laboratories (LNL) of the Italian National Institute for Nuclear Physics (INFN) in the framework of the INFN-Politecnico di Torino MESH Research Agreement. Work in Japan was partially supported by a Grant-in-Aid for Science Research (A) (Grant
No.17H01141) by the Japan Society for the Promotion of Science (JSPS).}

%

\end{document}